\documentclass[10pt,aps,twocolumn,prc,superscriptaddress,noshowpacs,nofootinbib,noshowkeys,floatfix]{revtex4}
\usepackage[dvips]{graphics,graphicx}
\usepackage[colorlinks=true,linktocpage=true,linkcolor=blue,citecolor=blue]{hyperref}
\usepackage[usenames,dvipsnames]{color}
\usepackage{amsmath, amssymb}
\usepackage{multirow}
\usepackage{longtable}
\usepackage{color}
\usepackage[normalem]{ulem}  

\renewcommand\sout{\bgroup \color{blue} \ULdepth=-.5ex \ULset}

\begin{document}

\preprint{}

\title{Effect of an anisotropic escape mechanism on elliptic flow in relativistic heavy-ion collisions}

\author{Amaresh Jaiswal}
\affiliation{School of Physical Sciences, National Institute of Science Education and Research, HBNI, Jatni-752050, India}
\author{Partha Pratim Bhaduri}
\affiliation{Variable Energy Cyclotron Centre, HBNI, 1/AF Bidhan Nagar, Kolkata 700 064, India}

\date{\today}

\begin{abstract}

We study the effect of anisotropic escape mechanism on elliptic flow in relativistic heavy-ion collisions. We use Glauber model to generate initial conditions and ignore hydrodynamic expansion in the transverse direction. We employ Beer-Lambert law to allow for the transmittance of produced hadrons in the medium and calculate the anisotropy generated due to the suppression of particles traversing through the medium. In order to separate non-flow contribution due to surface bias effects, we ignore hydrodynamic expansion in the transverse direction and consider purely longitudinal boost-invariant expansion. We calculate the transverse momentum dependence of elliptic flow, generated from anisotropic escape mechanism due to surface bias effects, for various centralities in $\sqrt{s_{NN}}=200$~GeV Au$+$Au collisions at RHIC and $\sqrt{s_{NN}}=2.76$~TeV Pb$+$Pb collisions at LHC. We find that the surface bias effects have a sizable contribution to the total elliptic flow observed in heavy-ion collisions indicating that the viscosity of the QCD matter extracted from hydrodynamic simulations may be underestimated.

\end{abstract}



\maketitle

\section{Introduction}

High energy heavy-ion collisions aim to create and study different phases of quantum chromodynamics (QCD) at extremely high temperature and density \cite{Braun-Munzinger:2015hba}. Observation of transverse flow, mass ordering and anisotropic flow in experimental data at the Relativistic Heavy Ion Collider (RHIC) and the Large Hadron Collider (LHC) indicates the presence of hydrodynamic phase in the evolution. Agreement of the experimental data with hydrodynamics inspired models suggest that the system is locally thermalized and possess a collective flow velocity due to the buildup of extremely high pressure which drives the system to expand at relativistic speed. Indeed relativistic hydrodynamics has been applied quite successfully to model high energy heavy-ion collisions \cite{Heinz:2013th, Gale:2013da, Jaiswal:2016hex, Florkowski:2017olj}. Moreover the application of relativistic viscous hydrodynamics to heavy-ion collisions has evoked widespread interest ever since the shear viscosity to entropy density ratio, $\eta/s$, estimated from the analysis of anisotropic flow data was found to be close to the conjectured lower bound of $1/4\pi$ \cite{Kovtun:2004de}. This led to the claim that the quark-gluon plasma (QGP) formed in relativistic heavy-ion collisions was the most perfect fluid ever observed.

No other theoretical construct has been able to describe the multitude of experimental data on heavy-ion collisions at the same level of accuracy as relativistic hydrodynamics \cite{Zhou:2015iba, Koop:2015wea, Bozek:2015swa, He:2015hfa, Romatschke:2015dha}. On the other hand, some theoretical ideas are able to mimic certain types of hydrodynamic signals \cite{Romatschke:2016vzy}. For instance, it was proposed that surface bias effects originating from the escape mechanism of the partons may have a large contribution to anisotropic flow even if the system interactions are small \cite{He:2015hfa}. This challenges the current understanding of the anisotropic flow generated from hydrodynamic pressure gradients. Moreover, if anisotropic flow originating from surface bias effects is indeed larger or even comparable to hydrodynamically driven collective flow, then the extracted $\eta/s$ is severely underestimated. Since escape is inevitable for a transient colliding system, it is imperative to examine the possible role of the escape mechanism on anisotropic flow within the framework of relativistic hydrodynamics. 

It is generally perceived that large elliptic anisotropy in momentum of the observed particles, also known as elliptic flow, can only be generated in heavy-ion collisions where the size of the medium is large enough to admit a hydrodynamic description. Of particular interest are non-central heavy-ion collisions where the overlap volume of the colliding nuclei is anisotropic in the transverse plane (perpendicular to beam). The pressure gradient due to initial geometrical anisotropy would generate anisotropic expansion and final-state elliptic flow. Indeed large elliptic flow has been measured in heavy-ion collisions at RHIC and LHC which can be well explained within the framework of relativistic hydrodynamics. However, recent particle correlation data hint at similar elliptic flow in high multiplicity $d\,+\,$Au collisions at RHIC \cite{Adare:2014keg} and, $p\,+\,p$ and $p\,+\,$Pb collisions at the LHC \cite{Abelev:2012ola, ABELEV:2013wsa, Abelev:2013haa, Abelev:2014mda, Sarkar:2016opn, Sarkar:2016ikv, Sarkar:2017rol, Ghosh:2014eqa, Kalaydzhyan:2015xba, Bautista:2015kwa, Gutay:2015cba}. While hydrodynamics seems to describe the experimental data well suggesting that these small-system collisions might create QGP, it is in contrast to general expectations based on the argument that equilibration can not be achieved in such small systems and therefore hydrodynamics might not be applicable. On the other hand, anisotropic escape mechanism due to surface bias effects might play a dominant role in generating large elliptic flow in these small-system collisions \cite{Koop:2015trj, Sarkar:2016nox}.

The consequences of a freeze-out criterion for heavy-ion collisions, based on pion escape probabilities from the rapidly expanding hot and dense medium, and the influence of expansion and scattering rate on the escape probability has been studied before in Ref.~\cite{Tomasik:2002qt}. Moreover, it was also found that a single collision per particle, in average, is already enough to generate sizable elliptic flow, with mass ordering between the species \cite{Borghini:2010hy}. This was indicative of the role of escape mechanism in mimicking certain type of hydrodynamic signals. In the context of heavy flavors, the propagation of charm and bottom quarks through an ellipsoidal QGP was studied in Refs.~\cite{Das:2010cz, Das:2011fe}. The elliptic flow thus obtained was due to difference in energy loss in the transverse plane originating from inequal path lengths in an ellipsoidal QGP. To this end, an experimentally accessible observable, which discriminates between collective and non-collective contributions to the observed elliptic flow, was also proposed in Ref.~\cite{Liao:2009ni}.

In this paper, we study the effect of escape mechanism due to shape anisotropy on elliptic flow in relativistic heavy-ion collisions. We use Glauber model to generate initial conditions and ignore hydrodynamic expansion in the transverse direction to separate out non-flow contribution due to surface bias effects. We employ Beer-Lambert law to allow for the transmittance of produced hadrons in the medium and calculate the anisotropy generated due to the suppression of particles traversing through the medium. While the analysis in Ref.~\cite{He:2015hfa} was performed within the framework of transport theory, we consider longitudinal boost-invariant hydrodynamic expansion in order to estimate the surface bias effects. We calculate the transverse momentum dependence of elliptic flow, generated from anisotropic escape mechanism due to surface bias effects, for various centralities in $\sqrt{s_{NN}}=200$~GeV Au$+$Au collisions at RHIC and $\sqrt{s_{NN}}=2.76$~TeV Pb$+$Pb collisions at LHC. We find that the surface bias effects have a sizable contribution to the total elliptic flow observed in these heavy-ion collisions. This indicates that that the viscosity of the QCD matter, extracted from hydrodynamic simulations, may be underestimated.


\section{The model}

We work in the Milne coordinate system ($\tau,x,y,\eta_s$), where $\tau=\sqrt{t^2-z^2}$ and $\eta_s=\tanh^{-1}(z/t)$. The metric tensor for this co-ordinate system is $g_{\mu\nu}={\rm diag}(1,\,-1,\,-1,\,-\tau^2)$. For longitudinal boost-invariant flow, i.e., $v^z=z/t$, the fluid four-velocity is given by $u^\mu=(1,0,0,0)$.  

In order to separate the surface bias effects due to source shape anisotropy on transverse momentum anisotropy of the observed particles, we assume vanishing fluid velocity in the transverse direction. This implies that the transverse energy density distribution of the system retains the initial shape and therefore all the momentum anisotropy would stem from the anisotropic escape mechanism due to surface bias effects. The time evolution of energy density of the system is therefore governed only by the Bjorken's scaling solution \cite{Bjorken:1982qr},
\begin{equation}\label{bjorken_exp}
\epsilon\propto\tau^{-4/3}.
\end{equation}
Using the above equation, the freeze-out time $\tau_f(x,y)$ can be obtained as a function of the initial energy density,
\begin{equation}\label{tauf_xy}
\tau_f(x,y) = \tau_i \left[ \frac{\epsilon_i(x,y)}{\epsilon_f} \right]^{3/4},
\end{equation}
where $\tau_i$ is the initialization time, $\epsilon_f$ is the freeze-out energy density and $\epsilon_i(x,y)$ initial energy density distribution obtained using the optical Glauber model of nuclear collisions.

The emitted hadron spectra can be obtained using the Cooper-Frye prescription for particle production \cite{Cooper:1974mv}
\begin{equation}\label{CF}
\frac{dN}{d^2p_Tdy} = \frac{g}{(2\pi)^3}\int p_\mu d\Sigma^\mu f(x,p),
\end{equation}
where $g$ is the degeneracy factor, $d\Sigma_\mu$ is the oriented freeze-out hyper-surface and $f(x,p)$ is the phase-space distribution function of the particles at freeze-out. In the present case we use classical Maxwell-Boltzmann distribution function for simplicity, $f_0=\exp(-u_\mu p^\mu/T)$. The above prescription should be modified to account for the re-absorption of the hadrons in the medium due to surface bias effects. We propose the following modification
\begin{equation}\label{CF_tr}
\frac{dN^T}{d^2p_Tdy} = \frac{g}{(2\pi)^3}\int p_\mu d\Sigma^\mu\, \Theta(x,p)\, f(x,p),
\end{equation}
where $\Theta(x,p)$ is the transmittance coefficient which can, in general, depend on the position and momentum of the produced hadrons.

The components of particle four-momenta, $p^\mu$ are given by
\begin{align}
p^\tau &= m_T \cosh(y-\eta_s), \nonumber\\
p^x &= p_T \cos\phi, \quad
p^y = p_T\sin\phi, \nonumber\\
p^{\eta_s} &= m_T \sinh(y-\eta_s)/\tau,
\end{align}
where $m_T^2=p_T^2+m^2$, $p_T$ is the transverse momentum, $y$ is the particle rapidity, and $\phi$ is the azimuthal angle in the momentum space. The freeze-out hyper-surface can be written as $d\Sigma_\mu = (m_T \cosh\eta_s, -\partial\tau_f/\partial x, -\partial\tau_f/\partial y, m_T \sinh\eta_s) \tau_f d\eta_s dx dy$, where $\tau_f(x,y)$ is the freeze-out time. Moreover, we note that in the absence of transverse expansion, $u\cdot p = m_T \cosh(y-\eta_s)$ and 
\begin{align}\label{hypersurface}
p^\mu d\Sigma_\mu = &\left[ m_T \cosh(y-\eta_s) - \left( p^x\frac{\partial\tau_f}{\partial x}
+ p^y\frac{\partial\tau_f}{\partial y} \right) \right] \nonumber\\
&\times\tau_f\, d\eta_s\, dx\, dy.
\end{align}
In the above equation, the derivatives of $\tau_f(x,y)$ indicates the curvature of the constant temperature freeze-out hyper-surface.

For longitudinal boost invariant flow where the transverse expansion is ignored, the spectra of emitted particles is
\begin{align}\label{emit_had}
\frac{dN}{d^2p_Tdy} =&\ \frac{g}{4 \pi^3}\Bigg[ m_T\, K_1 \int \tau_f\, dx\, dy \nonumber\\
&- K_0 \int \left( p^x\frac{\partial\tau_f}{\partial x}
+ p^y\frac{\partial\tau_f}{\partial y} \right)\tau_f\, dx\, dy \Bigg]
\end{align}
where $K_n \equiv K_n(z_m)$ are the modified Bessel functions of the second kind of order $n$ with argument $z_m\equiv m_T/T_f$ and $T_f$ is the freeze-out temperature corresponding to the freeze-out energy density $\epsilon_f$. Note that for the spectra of emitted hadrons given in Eq.~(\ref{emit_had}), the anisotropic flow defined as
\begin{equation}\label{vn}
v_n(p_T) \equiv 
\dfrac{\displaystyle{\int_{-\pi}^{\pi}}d\phi\,\cos(n\phi)\,\dfrac{dN}{dy\,p_T\,dp_T\,d\phi}}
{\displaystyle{\int_{-\pi}^{\pi}}d\phi\,\dfrac{dN}{dy\,p_T\,dp_T\,d\phi}},
\end{equation}
vanishes.

The hadrons produced using Eq.~(\ref{emit_had}) are emitted isotropically in $\phi$. However, it is important to note that the hadrons that have to traverse through the medium may be reabsorbed. Therefore one should take into account the probability of absorption of the hadrons in the medium. The probability of escape of the produced hadrons is given by the Beer-Lambert law for transmittance, $P_{\rm esc}=\exp(-\int\rho\,\sigma\,dl)$, where $\rho$ is the space-time dependent density of the system and $\sigma$ is interaction cross-section. One should keep in mind that the transmittance probability involves the entire future density evolution of the system, and is not captured by the Cooper–Frye freeze-out prescription.

In the present work, we use a modified version of the Beer-Lambert law for transmittance. We consider energy-density dependent attenuation of the hadrons traveling through the medium in the transverse plane. The transmittance probability is given by
\begin{equation}\label{transmittance}
\Theta(x,y,p_T,\phi) = \exp\left[-\int_0^\infty\frac{\sigma(p_T)}{T}\,\epsilon\left(x',\, y';\;\tau'\right)dl\right],
\end{equation}
where $x'=x+l\cos\phi$, $y'=y+l\sin\phi$ and $\tau'$ is the time at which the hadrons reaches the point $(x',y')$ in the transverse plane. In the above equation, we consider constant cross-section for the hadrons interacting with the medium
\begin{equation}\label{cross-section}
\sigma = \frac{T}{\epsilon_f\,\Lambda},
\end{equation}
where $\Lambda$ is the length scale associated with absorption of the hadrons inside the medium. The above form of $\sigma$ is motivated by transport calculations where one assumes constant cross-sections for the partonic and hadronic interactions.

Next we consider the evolution of the medium while the hadrons are traversing through it. The time at which the hadrons are produced is $\tau_f$ and the time at which they traverse a length $l$ to reach the point $(x',y')$ is 
\begin{equation}\label{adiab_ass_relx}
\tau' = \tau_f + \frac{l}{v_T}, \quad {\rm where} ~~ v_T = \frac{p_T}{m_T}.
\end{equation}
Using the Bjorken scaling. Eq.~(\ref{bjorken_exp}), we can obtain 
the energy density at time $\tau'$ to be
\begin{equation}\label{evol_eng_den}
\epsilon(x',\, y';\,\tau') = \epsilon(x',\, y';\,\tau_f) \left( 1 + \frac{l\,m_T}{\tau_f\,p_T} \right)^{-4/3}.
\end{equation}
Using the Bjorken scaling relation again, one can obtain $\epsilon(x',\, y';\,\tau_f)$ in terms of the initial energy density distribution,
\begin{equation}\label{en_den_tran}
\epsilon\left( x',\, y';\,\tau_f \right) 
= \epsilon_i\left( x',\, y' \right)\left[\frac{\tau_i}{\tau_f(x,y)} \right]^{4/3}
= \frac{\epsilon_i\left( x',\, y' \right)}{\epsilon_i\left( x,\, y \right)} \epsilon_f,
\end{equation}
where we have used Eq.~(\ref{tauf_xy}) to write the last equality.

Using Eqs.~(\ref{cross-section}), (\ref{evol_eng_den}) and (\ref{en_den_tran}), we see that Eq.~(\ref{transmittance}) can be written as
\begin{align}\label{transmittance_evol}
&\Theta(x,y,p_T,\phi) = \exp\Bigg[-\frac{1}{\epsilon_i\left( x,\, y \right)\,\Lambda}\int_0^\infty dl\; \nonumber\\
&\times \epsilon_i\left( x+l\cos\phi,\, y+l\sin\phi \right) 
\left( 1 + \frac{l\,m_T}{\tau_f\,p_T} \right)^{-4/3} \Bigg].
\end{align}
This is the final form of the transmittance coefficient which can be used in Eq.~(\ref{CF_tr}) to calculate the transmitted spectra of hadrons. 

The spectra of transmitted hadrons, obtained after Cooper-Frye freeze-out in Eq.~(\ref{emit_had}) with the transmittance probability proposed in Eq.~(\ref{CF_tr}), can be written as
\begin{align}\label{trans_had}
&\frac{dN^T}{d^2p_Tdy} =\ \frac{g}{4 \pi^3}\Bigg[ m_T\, K_1 \int \Theta(x,y,p_T,\phi)\, \tau_f\, dx\, dy \nonumber\\
&- K_0 \int \Theta(x,y,p_T,\phi) \left( p^x\frac{\partial\tau_f}{\partial x}
+ p^y\frac{\partial\tau_f}{\partial y} \right)\tau_f\, dx\, dy \Bigg].
\end{align}
Note that the spectra of transmitted hadrons now has $\phi$ dependence via the transmittance coefficient $\Theta$ and can lead to sizable anisotropic flow,
\begin{equation}\label{vn_trans}
v_n(p_T) \equiv 
\dfrac{\displaystyle{\int_{-\pi}^{\pi}}d\phi\,\cos(n\phi)\,\dfrac{dN^T}{dy\,p_T\,dp_T\,d\phi}}
{\displaystyle{\int_{-\pi}^{\pi}}d\phi\,\dfrac{dN^T}{dy\,p_T\,dp_T\,d\phi}}.
\end{equation}
In the present work, we focus only on the second harmonic of the anisotropic flow, $v_2(p_T)$, also known as the elliptic flow. This is the most dominant flow harmonic for non-central collisions and depends predominantly on the initial geometrical shape of the medium.

Here we point out some limitations of our model. We neglect the energy deposited in the medium by the absorbed hadrons and therefore the change in energy density of the medium. Since the hadrons are soft probes, the heating of the medium due to deposited energy can be safely neglected compared to cooling due to Bjorken expansion, and therefore not important for the current study. Another assumption is that we do not account for the change in the momentum of the emitted hadrons i.e., we assume that the hadrons are either absorbed or transmitted completely without energy loss. While it is difficult to account for the momentum loss of the hadrons, the uncertainty due to this on $v_2$ should not be large. We leave this study for future work.


\section{Results and discussions}

\begin{figure}[t]
\begin{center}
\includegraphics[width=\linewidth]{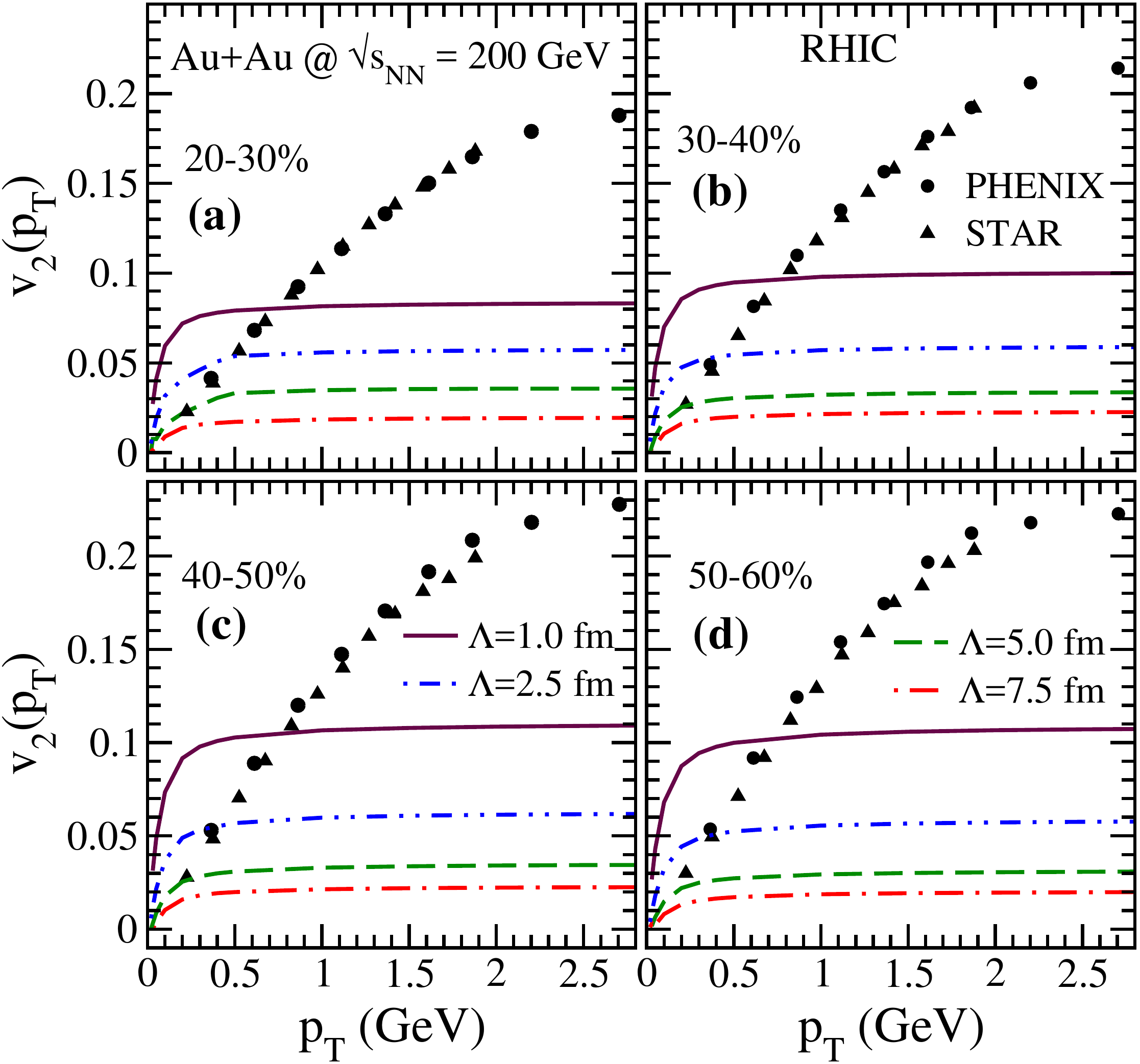}
\end{center}
\vspace{-0.8cm}
\caption{Transverse momentum dependence of elliptic flow of charged hadrons for RHIC at various centralities. The symbols represent experimental results from PHENIX and STAR collaborations. The theoritical curves are obtained by considering various values of attenuation length of hadrons in the medium. The error bars for the experimental data are contained within the symbol size.} 
\label{v2pT_RHIC}
\end{figure}

\begin{figure}[t]
\begin{center}
\includegraphics[width=\linewidth]{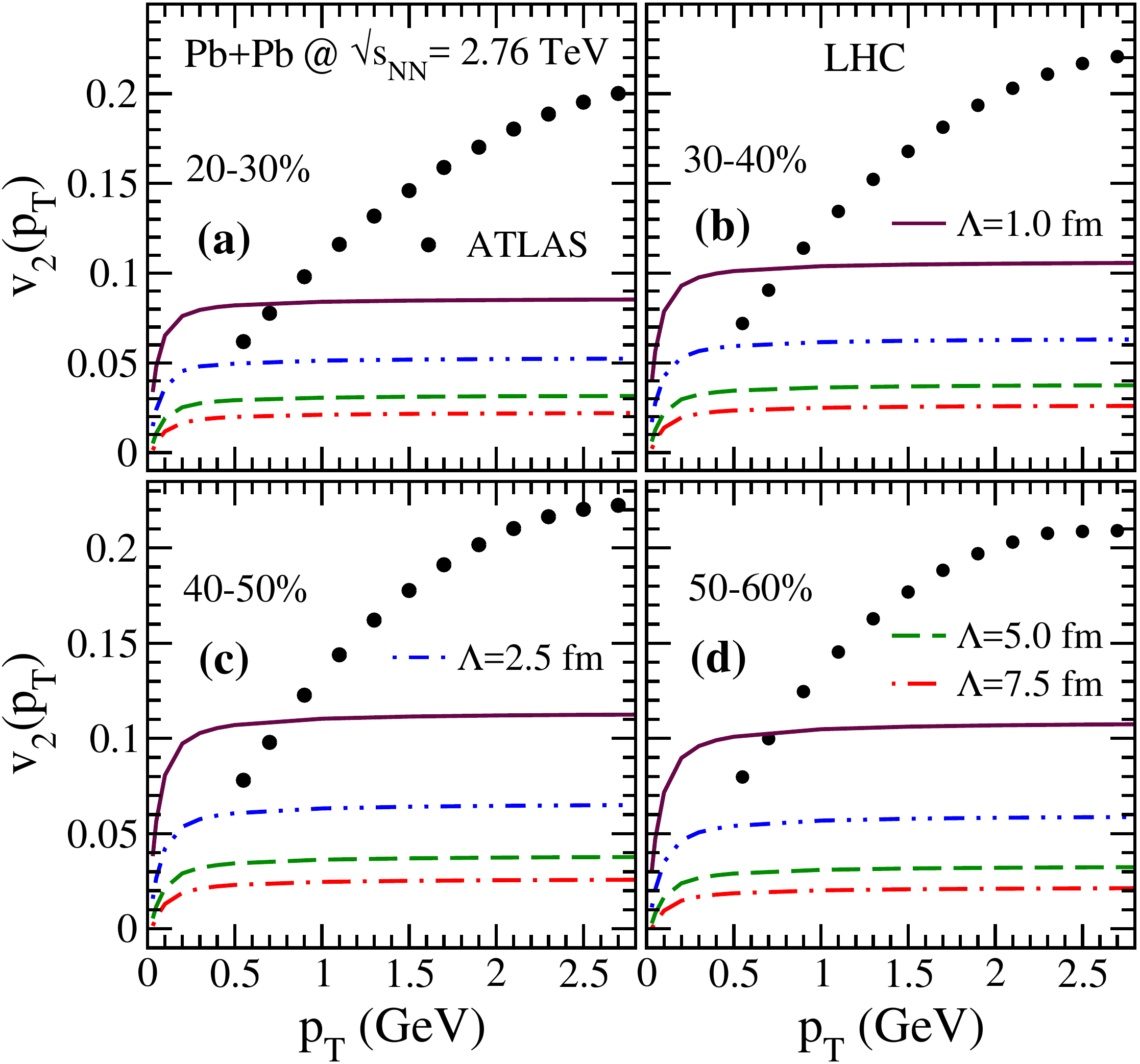}
\end{center}
\vspace{-0.8cm}
\caption{Transverse momentum dependence of elliptic flow of charged hadrons for LHC at various centralities. The symbols represent experimental results from ATLAS collaborations. The theoritical curves are obtained by considering various values of attenuation length of hadrons in the medium. The error bars for the experimental data are contained within the symbol size.} 
\label{v2pT_LHC}
\end{figure}

As a demonstration of our model, we apply it to calculate elliptic flow of charged hadrons for Au$+$Au collisions at $\sqrt{s_{NN}} = 200$ GeV at RHIC and Pb$+$Pb collisions at $\sqrt{s_{NN}} = 2.76$ TeV at LHC. In Fig.~\ref{v2pT_RHIC}, we show $p_T$ dependence of $v_2$, obtained form  Eq.~(\ref{vn_trans}), for various centralities at RHIC. The symbols in Fig.~\ref{v2pT_RHIC} are experimental results obtained by the PHENIX and STAR collaborations \cite{Adare:2011tg, Adams:2004bi, Adamczyk:2013waa}. Theoretical curves are generated for different values of $\Lambda=1,~2.5,~5$ and $7.5$ fm. For RHIC, the initial energy density at the center of the fireball is set to $\epsilon_i(0,0) = 30$ GeV$/$fm$^3$ and the initial thermalization time is taken as $\tau_i=0.6$ fm$/c$ \cite{Adler:2004zn}. We consider freeze-out at energy density corresponding to a temperature of $T_f=130$ MeV. The results for Pb$+$Pb collisions at LHC at $\sqrt{s_{NN}} = 2.76$ TeV are depicted in Fig.~\ref{v2pT_LHC} for similar values of $\Lambda$. The symbols in Fig.~\ref{v2pT_LHC} represents experimental results obtained by the ATLAS collaboration \cite{ATLAS:2012at}. The other parameters for LHC are taken as $\epsilon_i(0,0)=85$ GeV$/$fm$^3$, and $\tau_i=0.4$ fm$/c$ \cite{Abelev:2013vea}. The freeze-out energy density is taken to be the same as that at RHIC. 

In Figs.~\ref{v2pT_RHIC} and \ref{v2pT_LHC}, we observe that for very small transverse momentum, $p_T\leq 0.1$~GeV, the anisotropic escape mechanism does not generate any elliptic flow. This may be attributed to the fact that for very low energy hadrons, the medium expands faster than their propagation speed and the hadrons do not experience the shape anisotropy of the fireball. On the other hand, for hadrons above a certain transverse momentum, they are attenuated by a constant factor depending on the interaction cross-section and hence the absorption length scale, $\Lambda$, in the medium. For small value of $\Lambda$, the attenuation is larger and therefore larger elliptic flow is generated. However, we observe that, even for a very small attenuation length, $\Lambda=1$ fm (solid brown curves), the model is unable to generate the magnitude of experimentally observed elliptic flow. This shows that anisotropic escape is not sufficient to explain the experimental data and one needs hydrodynamic buildup of collective flow to get the correct magnitude of the observed elliptic flow. On the other hand, we see that the elliptic flow generated from anisotropic escape is non-negligible and should be corrected for in the Cooper-Frye prescription for freeze-out.

It is interesting to note that the $p_T$ dependence of $v_2$ in Figs.~\ref{v2pT_RHIC} and \ref{v2pT_LHC} is similar to that obtained in Ref.~\cite{Borghini:2010hy} where it was shown that, within a Boltzmann kinetic approach, a single collision per particle is enough to generate sizable elliptic flow. On the other hand, we find that, within the present model, escape mechanism alone is not enough to generate the observed elliptic flow in non-central heavy ion collisions. This is contrary to the findings of Ref.~\citep{He:2015hfa} where the authors claim that the majority of anisotropic flow is generated from escape mechanism within a Boltzmann transport approach. This may be attributed to the fact that while we consider an equilibrated system, the parton-parton interactions in the transport calculations may not be enough for the system to attain equilibrium. However, when the parton-parton interaction cross-section is increased, the authors of Ref.~\citep{He:2015hfa} find that the hydrodynamic-type contribution to the anisotropic flow starts to dominate over the contribution due to escape mechanism, which is in agreement with our findings. Therefore, we provide an extremely simple model to account for the escape mechanism which can easily be implemented in the Cooper-Frye freeze-out prescription in a realistic hydrodynamic calculation.


\section{Summary and conclusion}

In this paper, we have studied the effect of anisotropic escape mechanism due to source shape anisotropy on elliptic flow in relativistic heavy-ion collisions. We have used optical Glauber model to generate initial conditions and ignored hydrodynamic expansion in the transverse direction to separate out non-flow contributions from surface bias effects. In order to calculate the probability of transmittance of produced hadrons through the medium, we have employed Beer-Lambert law. We found that this method, to account for loss of hadrons inside the medium, leads to final momentum anisotropy of the observed hadrons even though the transverse fluid velocity vanishes. As a demonstration, we calculated the transverse momentum dependence of elliptic flow, generated from anisotropic escape mechanism due to surface bias effects, for various centralities in $\sqrt{s_{NN}}=200$~GeV Au$+$Au collisions at RHIC and $\sqrt{s_{NN}}=2.76$~TeV Pb$+$Pb collisions at LHC. We found that the surface bias effects have a sizable contribution to the total elliptic flow observed in these collisions. On the other hand, the elliptic flow generated from escape mechanism alone is not sufficient to explain the experimental data. This indicates that while hydrodynamic buildup of fluid velocity is necessary to get the correct magnitude of elliptic flow, the viscosity of the QCD matter extracted from hydrodynamic simulations may be underestimated.

At this juncture, it is important to point out that in the presence of transverse fluid velocity, the number of hadrons entering the medium is relatively less as the momentum of the hadrons is biased towards the outward direction. On the other hand, this effect becomes more important when one has fluctuations in the initial condition leading to irregularities in the freeze-out hyper-surface. This leads to negative Cooper-Frye contributions \cite{Oliinychenko:2014tqa} and hence one should correctly account for the suppression of these hadrons in a realistic hydrodynamic calculation after freeze-out. While this is left for future work, we note that the escape mechanism might play a dominant role in small systems where the formation of a hydrodynamic medium is questionable. Therefore, it will be interesting to explore the applicability of the present model to generate flow in small systems.


\begin{acknowledgments}

The authors would like to thank Volker Koch for useful discussions during the initial stages of this work. We thank Victor Roy and Debojit Sarkar for helpful comments. The authors gratefully acknowledge support from GSI Darmstadt where this project was initiated. A.J. is supported in part by the DST-INSPIRE faculty award under Grant No. DST/INSPIRE/04/2017/000038.

\end{acknowledgments}


\end{document}